\pdfoutput=1
\documentclass[twocolumn,showpacs,preprintnumbers]{revtex4}%
\usepackage{amsmath}
\usepackage{graphicx}
\usepackage{dcolumn}
\usepackage{bm}
\usepackage[latin1]{inputenc}
\usepackage{amsfonts}
\usepackage{amssymb}
\usepackage{appendix}%
\begin{document}
\title{Decoherence versus disentanglement for two qubits in a squeezed bath.}
\author{Maritza Hernandez, Miguel Orszag}
\address{Facultad de Física, Pontificia Universidad Católica de
Chile, Casilla 306, Santiago,Chile}

\begin{abstract}
We study the relation between the sudden death and revival of the entanglement
of two qubits in a common squeezed reservoir, and the normal decoherence, by
getting closer to the Decoherence Free Subspace and calculating the effect on
the death and revival times.

\end{abstract}

\maketitle

\section{\bigskip Introduction}

In one-party quantum systems, coherence is destroyed by the action of the
environment, a phenomena that is local and occurs asymptotically in time.

On the other hand, there could be multiparty systems, with non-local quantum
correlations, often referred to as quantum entanglement. The non-locality and
coherence of the quantum entangled states makes them very important in
applications such as quantum teleportation \cite{t}, quantum cryptography
\cite{cr}, dense coding \cite{c} , etc.

If the environment would act on the various parties the same way it acts on
single systems, one would expect that a measure of entanglement, say the
concurrence, would also decay exponentially in time.

However, this is not always the case. Recently Yu and Eberly
\cite{e1,e2,e3} showed that under certain conditions, the dynamics
could be completely different and the quantum entanglement may
vanish in a finite time. They called this effect "entanglement
sudden death". This effect has also been observed experimentally
\cite{d}.

A popular measure of entanglement is the concurrence \cite{WO}. For pure
states is defined as:
\begin{equation}
C(\Psi)=|\langle\Psi|\widetilde{\Psi}\rangle|,
\end{equation}
where:
\begin{equation}
|\widetilde{\Psi}\rangle_{12}=(\sigma_{y}^{1}\otimes\sigma_{y}^{2}%
)\otimes|\Psi\rangle_{12},\quad\text{for two qubits.}%
\end{equation}

For a mixed state, the concurrence is defined as:
\begin{equation}
C(\rho)=max\{0,\sqrt{\lambda_{1}}-\sqrt{\lambda_{2}}-\sqrt{\lambda_{3}}%
-\sqrt{\lambda_{4}}\},
\end{equation}
where the $\sqrt{\lambda_{i}}$ are the eigenvalues($\lambda_{1}$ being the
largest one)of a non-Hermitian matrix $\rho\widetilde{\rho}$, \ and
$\widetilde{\rho}$ is defined as:
\begin{equation}
\widetilde{\rho}=(\sigma_{y}^{1}\otimes\sigma_{y}^{2})\rho^{\ast}(\sigma
_{y}^{1}\otimes\sigma_{y}^{2}),
\end{equation}
$\rho^{\ast}$ being the complex conjugate of $\rho$.

In this paper we will explore the relation between the sudden death (and
revival) of the entanglement between the two two-level atoms in a squeezed
bath and the normal decoherence and the decoherence free subspace (DFS), which
in this case is a two-dimensional plane.

\section{The model}

In the present work, we consider two two-level atoms that interact
with a common squeezed reservoir, and we will focus on the
evolution of the entanglement between them, using as a basis, the
Decoherence Free Subspace states, as defined in references
\cite{MO} and  \cite{mo}.

The master equation, in the Interaction Picture, for a two-level system in a
broadband squeezed vacuum bath is given by \cite{MO2}:%

\begin{align}
\frac{\partial\rho}{\partial t}  &  =\frac{1}{2}\gamma(N+1)(2\sigma\rho
\sigma^{\dag}-\sigma^{\dag}\sigma\rho-\rho\sigma^{\dag}\sigma){}\nonumber\\
&  {}+\frac{1}{2}\gamma N(2\sigma^{\dag}\rho\sigma-\sigma\sigma^{\dag}%
\rho-\rho\sigma\sigma^{\dag}){}\nonumber\\
&  {}-\frac{1}{2}\gamma Me^{i\Psi}(2\sigma^{\dag}\rho\sigma^{\dag}%
-\sigma^{\dag}\sigma^{\dag}\rho-\rho\sigma^{\dag}\sigma^{\dag}){}\nonumber\\
&  {}-\frac{1}{2}\gamma Me^{i\Psi}(2\sigma\rho\sigma-\sigma\sigma\rho
-\rho\sigma\sigma),
\end{align}
where $\gamma$ is the spontaneous emission rate and $N=\sinh^{2}%
r,M=\sqrt{N(N+1)}$ and $\Psi$ are the squeeze parameters of the bath and
$\sigma^{\dagger},\sigma^{{}}$are the usual Pauli raising and lowering
matrices.\qquad\qquad\qquad\qquad\qquad\qquad\ \ \ \

It is simple to show that the above master equation can also be written in the
Lindblad form with a single Lindblad operator S%
\begin{equation}
\frac{\partial\rho}{\partial t}=\frac{1}{2}\gamma(2S\rho S^{\dag}-S^{\dag
}S\rho-\rho S^{\dag}S), \label{em}%
\end{equation}

with%
\[
S=\sqrt{N+1}(\sigma)-\sqrt{N}e^{i\Psi}(\sigma^{\dagger}){}.
\]
\qquad

For a two two-level system, the master equation has the same structure, but
now the S operator becomes:%
\begin{align}
S  &  =\sqrt{N+1}(\sigma_{1}+\sigma_{2})-\sqrt{N}e^{i\Psi}(\sigma_{1}^{^{\dag
}}+\sigma_{2}^{^{\dag}}){}\nonumber\\
&  =\cosh(r)(\sigma_{1}+\sigma_{2})-\sinh(r)e^{i\Psi}(\sigma_{1}^{^{\dag}%
}+\sigma_{2}^{^{\dag}}).
\end{align}

The Decoherence Free Subspace was found in ref \cite{MO} and consists of the
eigenstates of $S$ with zero eigenvalue. The states defined in this way, form
a two-dimensional plane in Hilbert Space. Two orthogonal vectors in this plane are:%

\begin{equation}
|\phi_{1} \rangle=\frac{1}{\sqrt{N^{2}+M^{2}}} (N |++ \rangle+ M e^{-i\Psi
}|--\rangle),
\end{equation}
\begin{equation}
|\phi_{2} \rangle=\frac{1}{\sqrt{2}} (|-+ \rangle- |+-\rangle).
\end{equation}

We can also define the states $|\phi_{3}\rangle$ and $|\phi_{4}\rangle$
orthogonal to the \{$|\phi_{1}\rangle$ , $|\phi_{2}\rangle\}$ plane:

\begin{equation}
|\phi_{3} \rangle=\frac{1}{\sqrt{2}} (|-+ \rangle+ |+-\rangle),
\end{equation}

\begin{equation}
|\phi_{4}\rangle=\frac{1}{\sqrt{N^{2}+M^{2}}}(M|++\rangle-Ne^{-i\Psi
}|--\rangle).
\end{equation}

To solve the master equation, we are going to use the basis $\{|\phi
_{1}\rangle,|\phi_{2}\rangle,|\phi_{3}\rangle,|\phi_{4}\rangle\}$ This
solution depends on the initial state. We present the general solution in the Appendix.

\bigskip In general, for density matrices written in the standard basis of the form:

\begin{equation}
\rho(t)=%
\begin{pmatrix}
\rho_{11} & 0 & 0 & \rho_{14}\\
0 & \rho_{22} & \rho_{23} & 0\\
0 & \rho_{32} & \rho_{33} & 0\\
\rho_{41} & 0 & 0 & \rho_{44}%
\end{pmatrix}
, \label{forma}%
\end{equation}

\bigskip one easily finds \cite{wang, zubairy} that the concurrence is given
by: $C(\rho)=max\{0,C1(\rho),C2(\rho)\}$, where: \bigskip%
\begin{align}
C1(\rho)  &  =2(\sqrt{\rho_{23}\rho_{32}}-\sqrt{\rho_{11}\rho_{44}%
})\label{concurrencia1}\\
C2(\rho)  &  =2(\sqrt{\rho_{14}\rho_{41}}-\sqrt{\rho_{22}\rho_{33}}).
\label{concurrencia2}%
\end{align}

\section{Solutions for initial states in DFS}

In this and the next two sections, all density matrices and expressions of
concurrences will be referred to the $\{|\phi_{1}\rangle,|\phi_{2}%
\rangle,|\phi_{3}\rangle,|\phi_{4}\rangle\}$ \ basis.

\begin{enumerate}
\item[a)] Consider $|\phi_{1}\rangle$ as the initial state. The solution of
master equation is given by:%
\begin{equation}
\rho(t)=%
\begin{pmatrix}
1 & 0 & 0 & 0\\
0 & 0 & 0 & 0\\
0 & 0 & 0 & 0\\
0 & 0 & 0 & 0
\end{pmatrix}
.
\end{equation}

This corresponds to an invariant state, and its concurrence is given
by:$\ C(\rho)=\frac{2\sqrt{N(N+1)}}{2N+1}$, which is a constant in time.

The concurrence only depends of \ $N$. For $N=0$ we have a factorized state at
all times, but as we increase $N$ ,\ we get a maximally entangled state in the
large N limit.(see, fig.\ref{c1}).%
\begin{figure}
[ptbh]
\begin{center}
\includegraphics[natheight=2.674000in,
natwidth=5.366200in, height=2.8383in, width=2.8383in
]%
{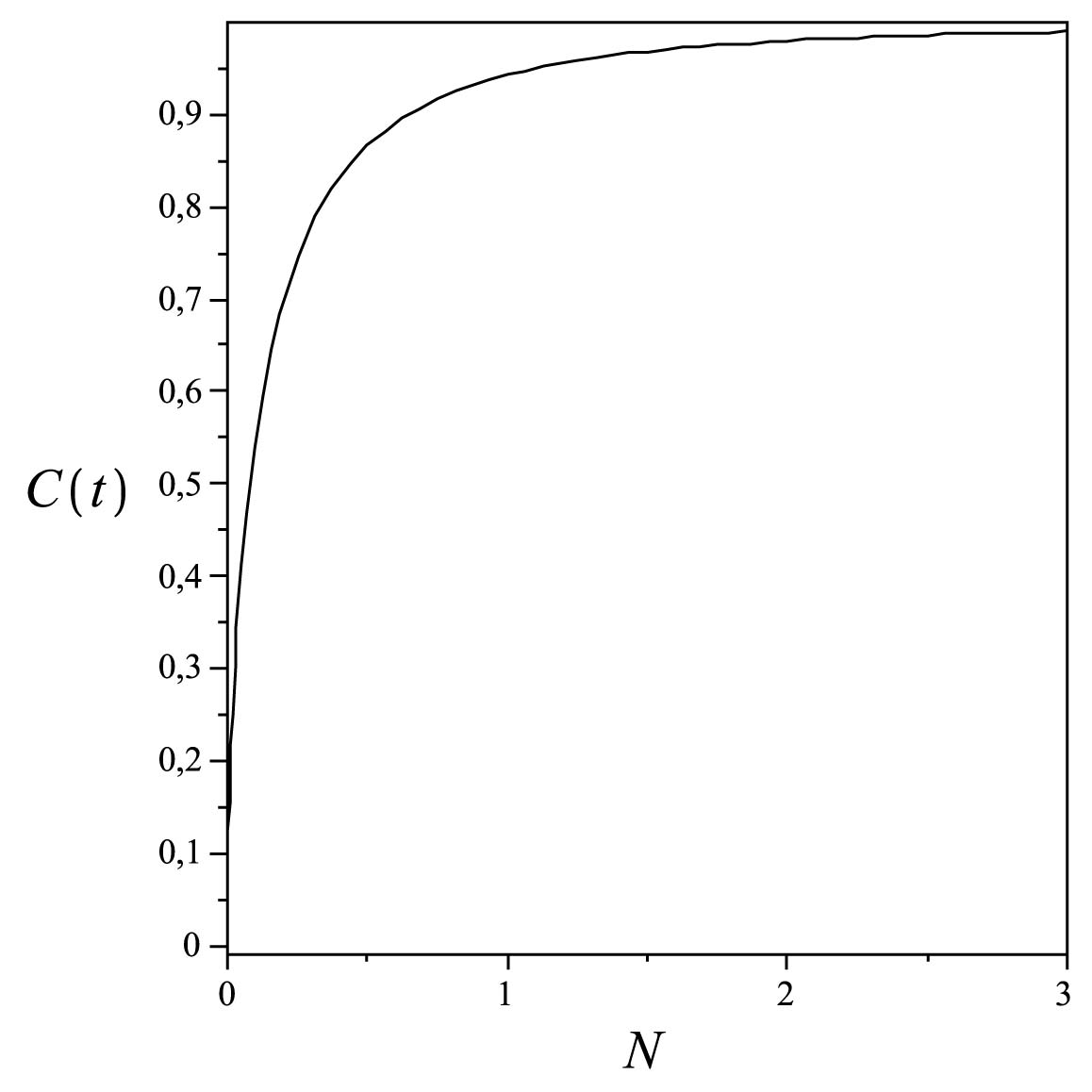}%
\caption{Concurrence as function of $N$, for $|\phi_{1}\rangle$ \ as the
initial state.}%
\label{c1}%
\end{center}
\end{figure}

\item[b)] If we now consider $|\phi_{2}\rangle$ as the initial state, we get
the solution of the master equation given by:%
\begin{equation}
\rho(t)=%
\begin{pmatrix}
0 & 0 & 0 & 0\\
0 & 1 & 0 & 0\\
0 & 0 & 0 & 0\\
0 & 0 & 0 & 0
\end{pmatrix}
.
\end{equation}
This state is also an invariant state and its concurrence is independent of
time: $C(\rho)=1.$
\end{enumerate}

\bigskip

In the following sections we consider, as initial states, $|\phi_{3}\rangle$
or $|\phi_{4}\rangle$ , and also superpositions of the form $\varepsilon$
$|\phi_{1}\rangle+\sqrt{1-\varepsilon^{2}}$ $|\phi_{4}\rangle$ and
$\varepsilon$ $|\phi_{2}\rangle+\sqrt{1-\varepsilon^{2}}$ $|\phi_{3}\rangle.$
The idea is to increase $\varepsilon$ and to study the effect of having an
increased component in the DFS on the death time of the entanglement. For
simplicity, we assume $\gamma=1$, and $\psi=0$.

\section{Solutions for $N=0$}

\begin{enumerate}
\item[a)] The third initial state considered is $|\phi_{3}\rangle$. Initially
its concurrence is: $C(\phi_{3}(0))=1.$ It \ corresponds to a maximally
entangled state.

The solution of master equation for this initial condition and $N=0$ is given
by:%
\begin{equation}
\rho(t)=%
\begin{pmatrix}
(e^{2t}-1)e^{-2t} & 0 & 0 & 0\\
0 & 0 & 0 & 0\\
0 & 0 & e^{-2t} & 0\\
0 & 0 & 0 & 0
\end{pmatrix}
.
\end{equation}

Since the matrix $\rho(t)\widetilde{\rho}(t)$ has only one nonzero eigenvalue,
in this case we use the separability criterion \cite{PERES}. According to this
criterion, the necessary condition for separability is that a matrix
$\rho^{PT}$, obtained by partial transposition of $\rho$, should have only
non-negative eigenvalues. In this particular case, we observe a negative
eigenvalue for all times, so the state stays entangled, (Fig.\ref{autovalor}).%
\begin{figure}
[ptbh]
\begin{center}
\includegraphics[natheight=2.674000in,
natwidth=5.366200in, height=2.3367in, width=2.3367in
]%
{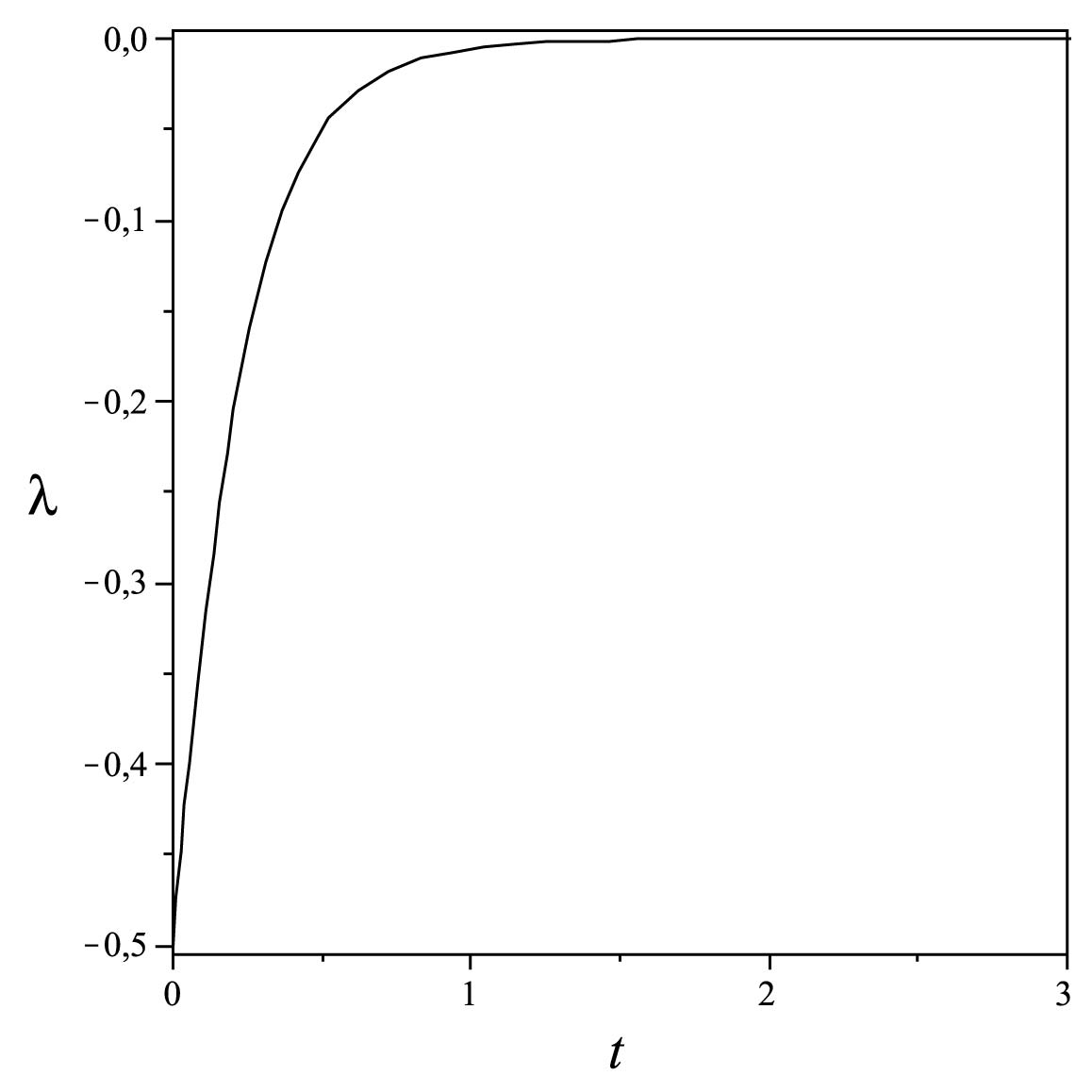}%
\caption{Negative eigenvalue $(\lambda)$ of $\rho^{PT}$, for the separability
criterion, for $|\phi_{3}\rangle$ as the initial state. This eigenvalue is
always negative , indicating entanglement at all times.}%
\label{autovalor}%
\end{center}
\end{figure}

\item[b)] Consider the initial state $|\phi_{4}\rangle$. Since $N=0$,
$|\phi_{4}\rangle=|++\rangle$ and $C(\phi_{4}(0))=0$, since is a factorized state.

The solution of master equation for this initial condition is given by:%
\begin{equation}
\rho(t)=%
\begin{pmatrix}
(-1-2t+e^{2t})e^{-2t} & 0 & 0 & 0\\
0 & 0 & 0 & 0\\
0 & 0 & 2te^{-2t} & 0\\
0 & 0 & 0 & e^{-2t}%
\end{pmatrix}
,
\end{equation}

and its concurrence is $C(\rho(t))=0$.

\item[c)] Now, we consider an initial superposition of $|\phi_{1}\rangle$ and
$|\phi_{4}\rangle$ : $|\Psi_{1}\rangle=\varepsilon$ $|\phi_{1}\rangle
+\sqrt{1-\varepsilon^{2}}$ $|\phi_{4}\rangle$.

As we increase $\varepsilon$, starting from $\varepsilon=0$, we
increase the initial projection onto the DFS. For $\varepsilon=1$
the initial state is in the DFS plane.

For $N=0$ \ we have: $|\Psi_{1}\rangle=\varepsilon$ $|--\rangle+\sqrt
{1-\varepsilon^{2}}$ $|++\rangle$. Its initial concurrence is $C(\Psi
_{1}(0))=2\varepsilon\sqrt{1-\varepsilon^{2}}$.

The solution of master equation for this initial condition with $N=0$ is given
by:%
\begin{equation}
\rho(t)=%
\begin{pmatrix}
\frac{(-2t-1+2t\varepsilon^{2}+\varepsilon^{2}+e^{2t})}{e^{2t}} & 0 & 0 &
\frac{\varepsilon\sqrt{1-\varepsilon^{2}}}{e^{t}}\\
0 & 0 & 0 & 0\\
0 & 0 & \frac{2t(1-\varepsilon^{2})}{e^{2t}} & 0\\
\frac{\varepsilon\sqrt{1-\varepsilon^{2}}}{e^{t}} & 0 & 0 & \frac
{(1-\varepsilon^{2})}{e^{2t}}%
\end{pmatrix}
,
\end{equation}

and the corresponding concurrence is given by:%
\begin{align}
C(\rho) &  =\max\{0,2\rho_{14}-\rho_{33}\}\\
&  =\max\{0,2((\varepsilon\sqrt{1-\varepsilon^{2}})e^{-t}-te^{-2t}%
(1-\varepsilon^{2}))\}
\end{align}
which is shown in fig.\ref{c2} for various values of $\varepsilon$:%
\begin{figure}
[ptbh]
\begin{center}
\includegraphics[natheight=2.674000in,
natwidth=5.366200in, height=2.3108in, width=2.3367in
]%
{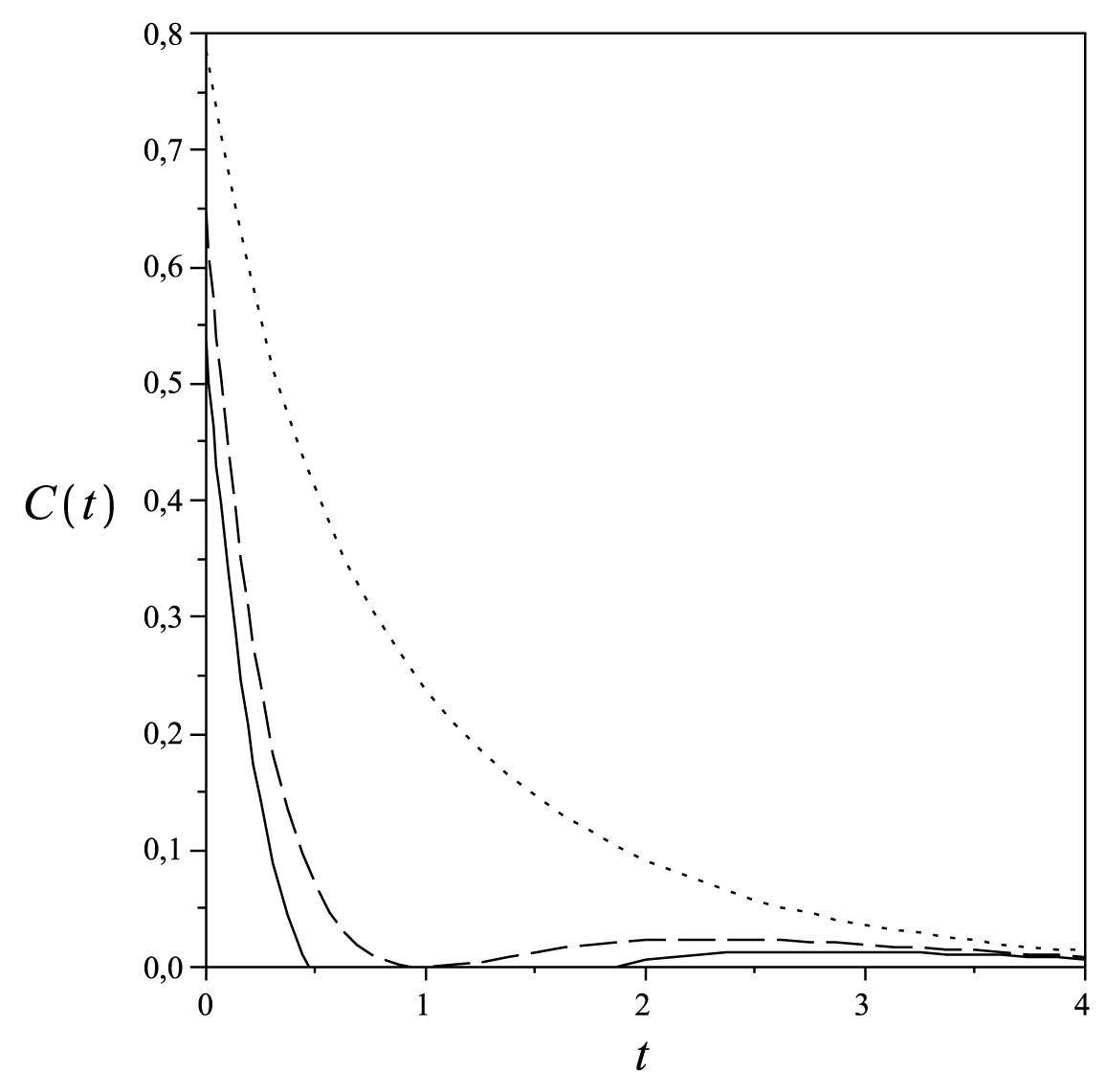}%
\caption{Time evolution of the concurrence for initial $|\Psi_{1}\rangle$ with
: $\varepsilon=0.28$ (solid line), $\varepsilon=0.345$ (dashed line),
$\varepsilon=0.9$ (dotted line),}%
\label{c2}%
\end{center}
\end{figure}

For $\varepsilon=0$ and $\varepsilon=1$ , its concurrence is zero, therefore
we have a non-entangled state. For $\varepsilon>0$ the initial entanglement
decreases in time, and the system becomes disentangled (sudden death) at a
time satisfying the relation:%
\begin{equation}
te^{-t}=\frac{\varepsilon}{\sqrt{1-\varepsilon^{2}}}, \label{Relacion}%
\end{equation}

For $0<\varepsilon<0.34525$ \ the equation (\ref{Relacion}) has two solutions,
namely, $t_{d}$ when the system becomes separable, and $t_{r}\geq t_{d}$ when
the entanglement revives. It should be noted that there is a critical
$\varepsilon$ for which $t_{d}=t_{r}$. For $\ 0.34525<\varepsilon<1$ the above
equation has no solution and the concurrence vanishes asymptotically in time.

So, when we are "not far" from $|\phi_{4}\rangle$ we observe a sudden death
\ and revival, but when we get "near" $|\phi_{1}\rangle$ \ this phenomenon
disappears. Fig.\ref{t1} shows the behavior of the death and revival time as
function of $\varepsilon.$
\begin{figure}
[ptbh]
\begin{center}
\includegraphics[
natheight=2.674000in, natwidth=5.366200in, height=1.4278in,
width=2.8383in
]%
{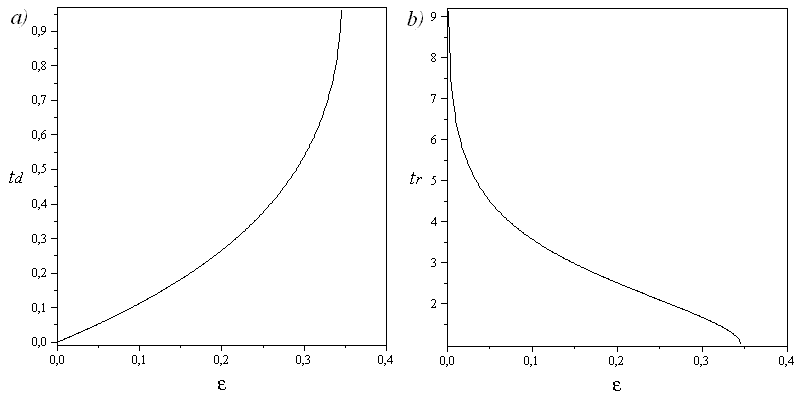}%
\caption{\textbf{a)} The death time \textbf{b)}The revival time of the
entanglement as a function of $\varepsilon$, with initial $|\Psi_{1}\rangle.$}%
\label{t1}%
\end{center}
\end{figure}
\ \ \ \ \ \ \ \ \ \ \ \ \ \ \ \ \ \ \ \ \ \ \ \ \ \ \ \ \ \ \ \ \ \ \ \ \ \ \ \ \ \ \ \ \ \ \ \ \ \ \ \ \ \ \ \ \ \ \ \ \ \ \ \ \ \ \ \ \ \ \ \ \ \ \ \ \ \ \ \ \ \ \ \

\item[d)] Finally, we consider an initial superposition of $|\phi_{2}\rangle$
and $|\phi_{3}\rangle$ : $|\Psi_{2}(0)\rangle=\varepsilon$ $|\phi_{2}%
\rangle+\sqrt{1-\varepsilon^{2}}$ $|\phi_{3}\rangle$, which is independent of
$N$. \ So, like in the pervious cases, as we increase $\varepsilon$, starting
from $\varepsilon=0$, we increase the initial projection onto the DFS. For
$\varepsilon=1$ the initial state is in the DFS plane.

For $N=0$ \ we have: $|\Psi_{2}\rangle=\frac{1}{\sqrt{2}}[(\varepsilon
+\sqrt{1-\varepsilon^{2}})$ $|-+\rangle-(\varepsilon-\sqrt{1-\varepsilon^{2}%
})$ $|+-\rangle]$. Its initial concurrence is $C(\Psi_{2}(0))=|2\varepsilon
^{2}-1|$.

The solution of master equation for this initial condition is given by:%
\begin{equation}
\rho(t)=%
\begin{pmatrix}
\frac{(e^{2t}-\varepsilon^{2}e^{2t}-1+\varepsilon^{2})}{e^{2t}} & 0 & 0 & 0\\
0 & \varepsilon^{2} & \frac{\varepsilon\sqrt{1-\varepsilon^{2}}}{e^{t}} & 0\\
0 & \frac{\varepsilon\sqrt{1-\varepsilon^{2}}}{e^{t}} & \frac{(1-\varepsilon
^{2})}{e^{2t}} & 0\\
0 & 0 & 0 & 0
\end{pmatrix}
,
\end{equation}

and the corresponding concurrence is:%
\begin{align}
C(\rho) &  =\max\{0,|\rho_{33}-\rho_{22}|\}\\
&  =\max\{0,e^{-2t}|\varepsilon^{2}e^{2t}-1+\varepsilon^{2}|\},
\end{align}
which is shown in fig.\ref{c3}.
\begin{figure}
[ptbh]
\begin{center}
\includegraphics[natheight=2.674000in,
natwidth=5.366200in, height=2.3367in, width=2.3367in
]%
{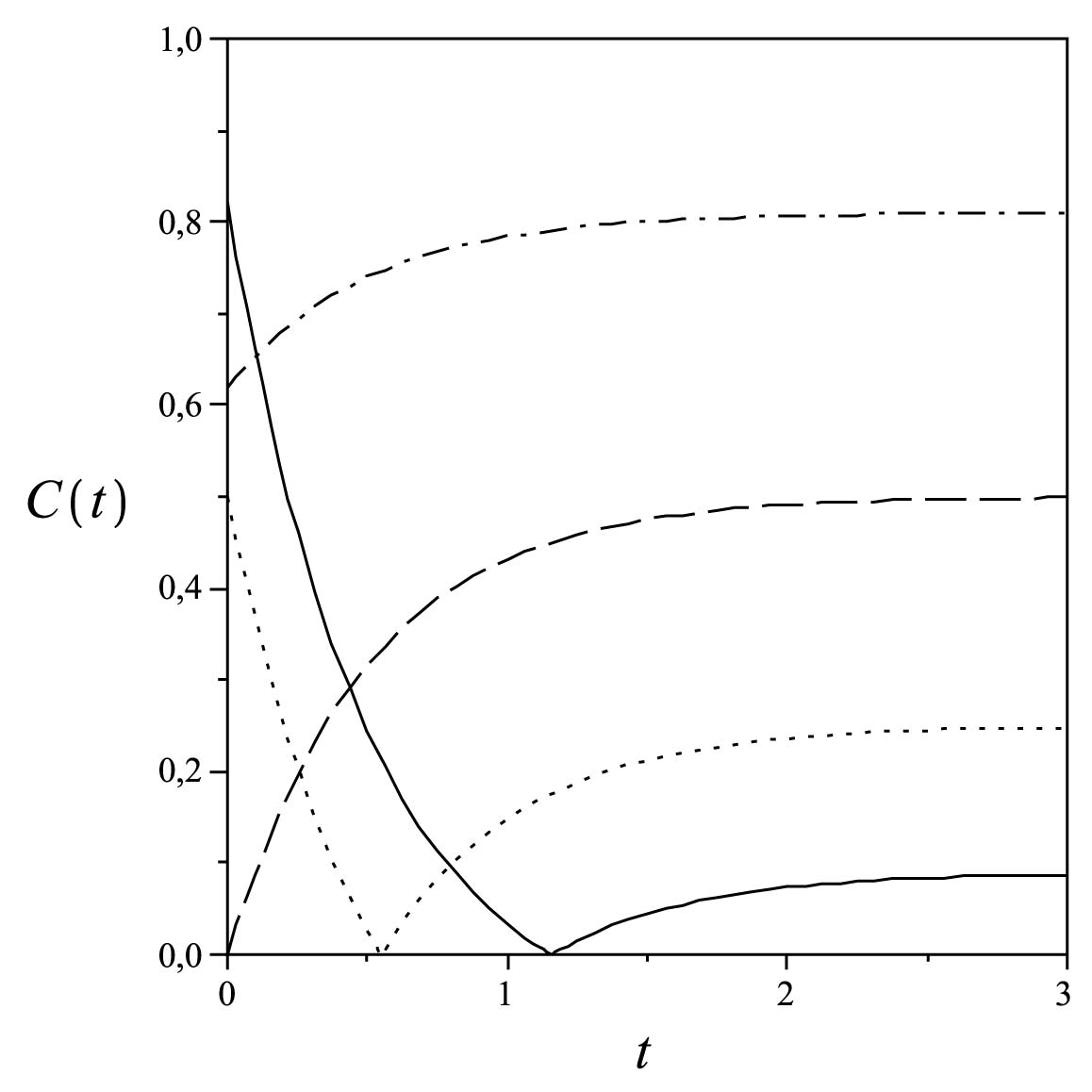}%
\caption{Time evolution of the concurrence with intial $|\Psi_{2}\rangle$ ,
for : $\varepsilon=0.3$ (solid line), $\varepsilon=0.5$ (dotted line),
$\varepsilon=0.707$ (dashed line), $\varepsilon=0.9$ (dash dotted line)}%
\label{c3}%
\end{center}
\end{figure}

For $0<\varepsilon<0.707$ the initial entanglement decreases in time, and the
system becomes disentangled instantaneously, at a time given by:%
\begin{equation}
t=\frac{1}{2}\ln(\frac{1-\varepsilon^{2}}{\varepsilon^{2}}). \label{tiempo}%
\end{equation}

However, at the same time, the entanglement revives reaching asymptotically
its stationary value.(fig. \ref{t2})%
\begin{figure}
[ptbh]
\begin{center}
\includegraphics[natheight=2.674000in,
natwidth=5.366200in, height=2.3367in, width=2.3367in
]%
{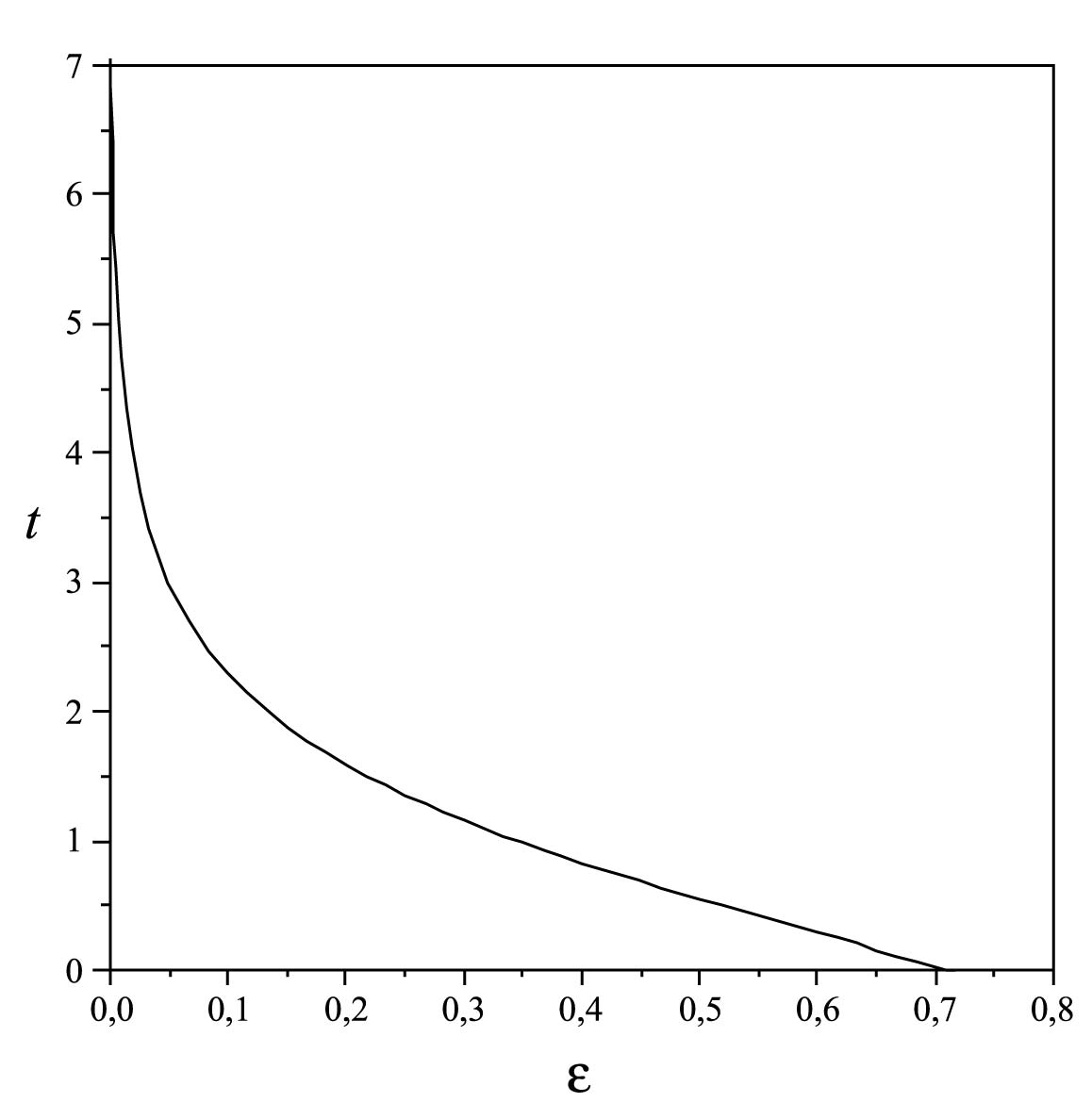}%
\caption{Death-revival time as given by eq \ref{tiempo} , versus
$\varepsilon.$}%
\label{t2}%
\end{center}
\end{figure}

When we approach the decoherence free subspace \ this phenomenon disappears.
\end{enumerate}

Next, we treat the cases with $N>0.$

\bigskip

\section{Solutions for $N\neq0$}

\bigskip In general the solution of master equation for $|\phi_{3}\rangle$ ,
$|\phi_{4}\rangle$ and $|\Psi_{1}\rangle$ \ as initial condition and $N\neq0$,
has the following form:%
\[
\rho(t)=%
\begin{pmatrix}
\rho_{11} & 0 & 0 & \rho_{14}\\
0 & 0 & 0 & 0\\
0 & 0 & \rho_{33} & 0\\
\rho_{41} & 0 & 0 & \rho_{44}%
\end{pmatrix}
,
\]
and written at the standard basis, has the same structure as in \ (\ref{forma}%
). Its concurrence is given by: $C(\rho)=max\{0,C1(\rho),C2(\rho)\}$, with the
explicit expressions for $C1(\rho)$ and $C2(\rho)$ are given by:%
\begin{align}
C1(\rho)  &  =2(\frac{\rho_{33}}{2}\nonumber\\
&  -\sqrt{\frac{(\rho_{11}N+\rho_{44}(N+1)+2\rho14\sqrt{N(N+1)})}{2N+1}%
}\nonumber\\
&  \times\sqrt{\frac{(\rho_{44}N+\rho_{11}(N+1)-2\rho14\sqrt{N(N+1)})}{2N+1}%
}),\\
C2(\rho)  &  =2(\frac{|\sqrt{N(N+1)}(\rho_{11}-\rho_{44})+\rho_{14}|}%
{2N+1}-\frac{\rho_{33}}{2}). \label{C2}%
\end{align}

On the other hand, for the initial condition $|\Psi_{2}\rangle$ , the
corresponding expressions for the density matrix and concurrence are:%
\[
\rho(t)=%
\begin{pmatrix}
\rho_{11} & 0 & 0 & \rho_{14}\\
0 & \rho_{22} & \rho_{23} & 0\\
0 & \rho_{32} & \rho_{33} & 0\\
\rho_{41} & 0 & 0 & \rho_{44}%
\end{pmatrix}
,
\]%
\begin{align}
C1(\rho)  &  =|\rho_{33}-\rho_{22}|\nonumber\\
&  -2\sqrt{\frac{N(\rho_{11}+\rho_{44})+\rho_{44}+2\rho_{14}\sqrt{N(N+1)}%
}{2N+1}}\nonumber\\
&  \times\sqrt{\frac{N(\rho_{11}+\rho_{44})+\rho_{11}-2\rho_{14}\sqrt{N(N+1)}%
}{2N+1}}, \label{C21}%
\end{align}%
\begin{align}
C2(\rho)  &  =\frac{2}{2N+1}|\sqrt{N(N+1)}(\rho_{11}-\rho_{44})+\rho
_{14}|\nonumber\\
&  -\sqrt{(\rho_{22}-2\rho_{23}+\rho_{33})(\rho_{22}+2\rho_{23}+\rho_{33})}
\label{C22}%
\end{align}

\begin{enumerate}
\item[a)] Next, we consider again the case for initial $|\phi_{3}\rangle$ ,
but for $N\neq0$. The concurrence is: $C(\rho)=max\{0,C1(\rho),C2(\rho)\}$,
with its initial value $C(\phi_{3}(0))=1$.

In fig.\ref{c4}, we show $C(t)$ versus time for various values of $N$. We
observe sudden death in a finite time, then the concurrence remains zero for a
period of time until \ the entanglement revives, and the concurrence reaches
asymptotically its stationary value. Notice that this time period increases
with $N.$%
\begin{figure}
[ptbh]
\begin{center}
\includegraphics[natheight=2.674000in,
natwidth=5.366200in, height=2.5374in, width=2.5374in
]%
{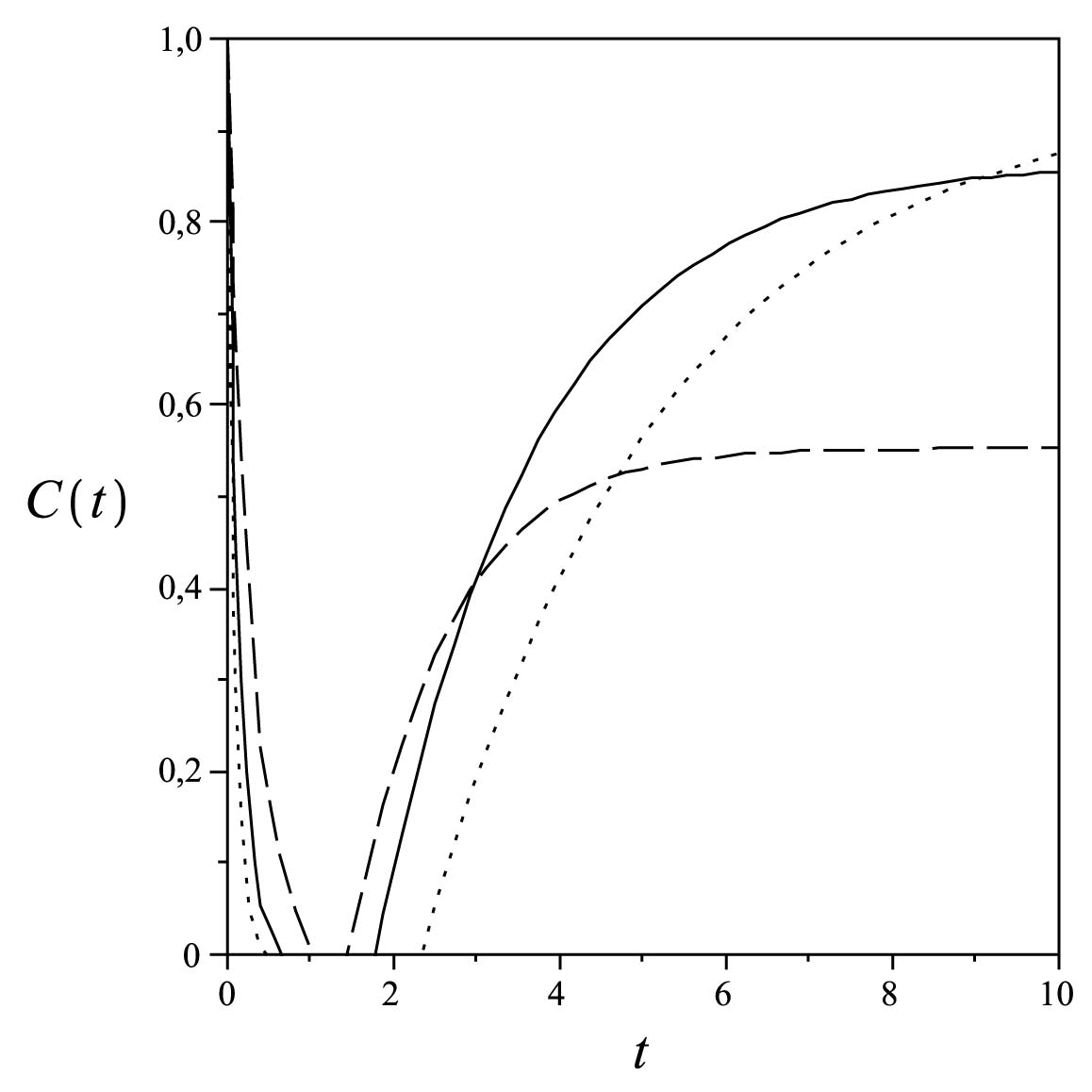}%
\caption{Time evolution of concurrence for initial $|\phi_{3}\rangle$ , with:
N=0.1 (dashed line), N=0.5 (solid line), N=1 (dotted line).}%
\label{c4}%
\end{center}
\end{figure}

In the fig.\ref{t3} we show the death and revival times versus N. They
decrease and increase with N respectively.
\begin{figure}
[ptbh]
\begin{center}
\includegraphics[
natheight=2.674000in, natwidth=5.366200in,
 height=1.26in,
width=2.53in
]%
{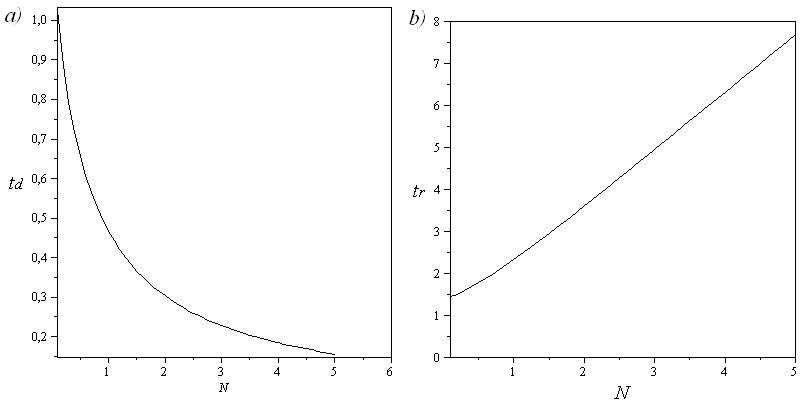}%
\caption{a)Death time b) Revival time versus N for the initial state $\mid
\phi_{3}\rangle.$}%
\label{t3}%
\end{center}
\end{figure}

\item[b)] Consider $|\phi_{4}\rangle$ as an initial state. The concurrence, in
this case, is: $C(\rho)=max\{0,C2(\rho)\}$, with $C2(\rho)$ defined by
(\ref{C2}). Initially, it takes the value $C(\phi_{4}(0))=\frac{2\sqrt
{N(N+1)}}{2N+1}.$ The behavior of concurrence is similar as in $|\phi
_{3}\rangle$. The initial entanglement quickly decays to zero, getting
disentanglement for a finite time interval , then the entanglement revives and
asymptotically it reaches its stationary value. \

However, unlike the case with initial state $|\phi_{3}\rangle$, the death time
first increases reaching a critical $N=0.421$, then decreases, as shown in Fig
\ref{t4}. The revival time has the same behavior as in $|\phi_{3}\rangle$.%
\begin{figure}
[ptbh]
\begin{center}
\includegraphics[
natheight=2.674000in, natwidth=5.366200in, height=1.5281in,
width=3.039in
]%
{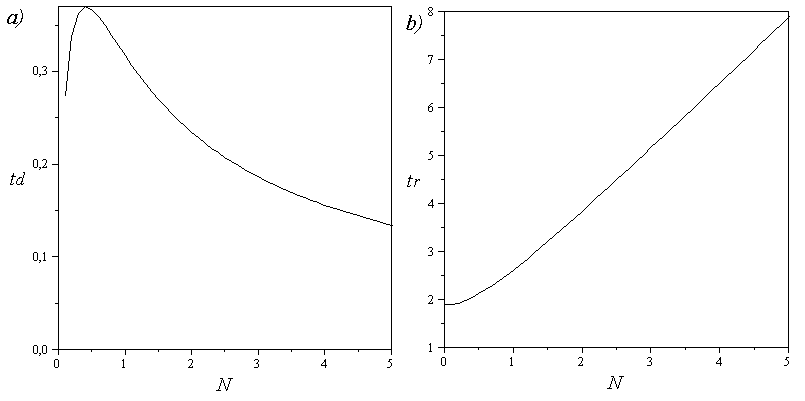}%
\caption{Death (a) and Revival (b) times versus N for the initial state
$\mid\phi_{4}\rangle.$}%
\label{t4}%
\end{center}
\end{figure}

\item[c)] In the following case, we consider the superposition $|\Psi
_{1}(t)\rangle=\varepsilon|\phi_{1}\rangle+\sqrt{1-\varepsilon^{2}}|\phi
_{4}\rangle$ as the initial state. The solution of master equation for this
initial condition depends on $\varepsilon$ and $N$ and also its concurrence,
which is: $C(\rho)=\max\{0,C2(t)\}$, where $C2(t)$ is in \ref{C2}. Since
$C1(t)$ is always negative, the only contribution to the concurrence comes
from $C2(t)$, . Its initial value being $C(\Psi_{1}(0))=\frac{|2\varepsilon
\sqrt{1-\varepsilon^{2}}+4\sqrt{N(N+1)}(\varepsilon^{2}-\frac{1}{2})|}{2N+1}$
, hence is it clear that for certain pairs of $N$ and $\varepsilon$, our
initial state will be a non-separable one. In the Fig. \ref{c6}\ \ we show the
time evolution of the concurrence for $N=0.1$ and several values of
$\varepsilon$. For $\varepsilon=0$ and $\varepsilon=1$ we retrieve $|\phi
_{4}\rangle$ and $|\phi_{1}\rangle$ respectively. For $0<\varepsilon<0.5$
\ the concurrence dies in a finite time, stays zero for a time interval and
subsequently revives, going asymptotically to its stationary value. For values
larger than $\varepsilon=0.5,$ there is no more sudden death, since we are
getting "close" to the DFS, and $C(t)$ goes asymptotically to its stationary
value.
\begin{figure}
[ptbh]
\begin{center}
\includegraphics[natheight=2.674000in,
natwidth=5.366200in, height=2.3367in, width=2.3367in
]%
{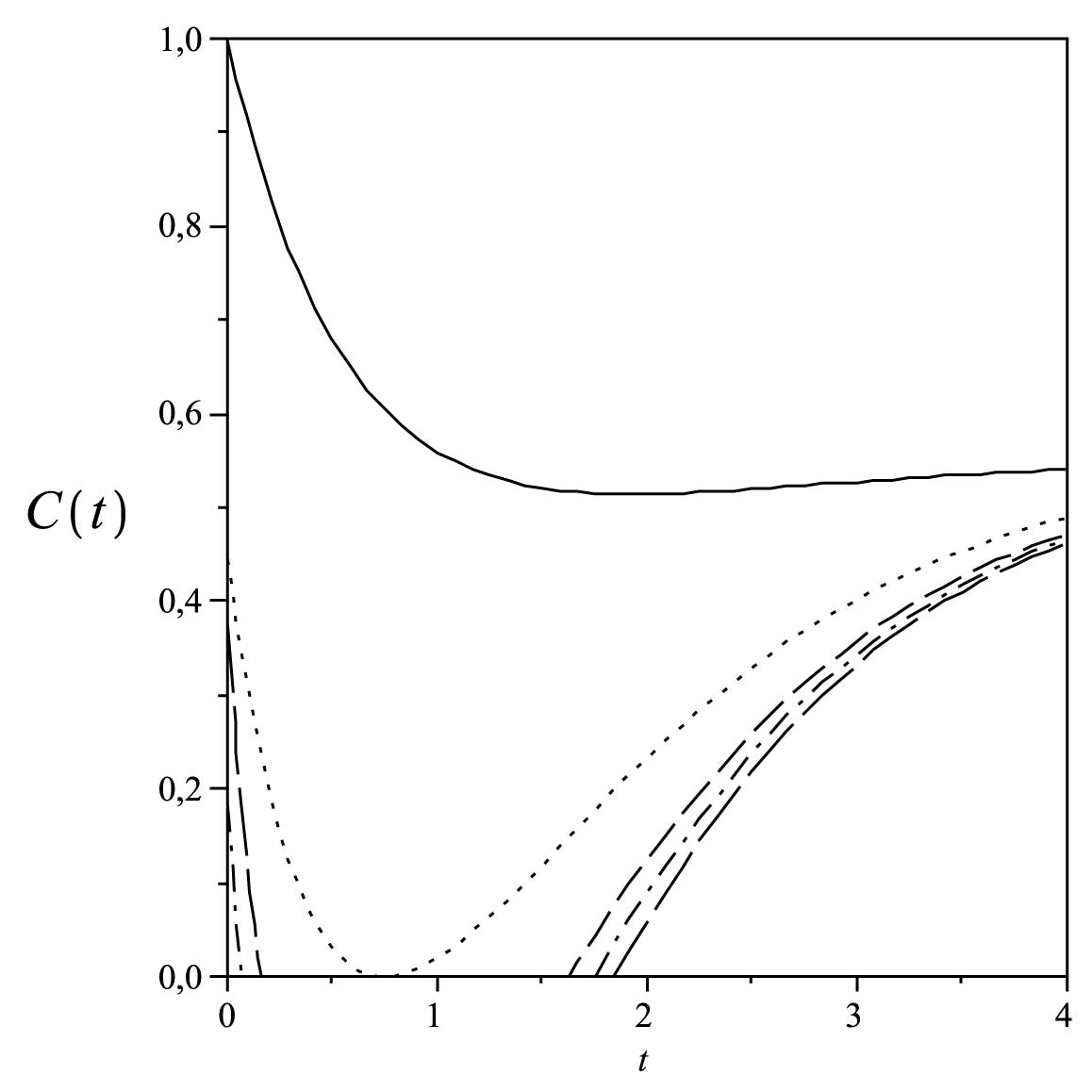}%
\caption{Time evolution of Concurrence for $|\Psi_{1}(t)\rangle$ as initial
state and $N=0.1$: $\varepsilon=0.1$ (long dashed line), $\varepsilon=0.2$
(dash dotted line), $\varepsilon=0.29$ (dashed line), $\varepsilon=0.5$
(dotted line), $\epsilon=0.9$ (solid line).}%
\label{c6}%
\end{center}
\end{figure}
\ The Figure \ref{t5} shows the death times versus $\varepsilon$ for
$N=\{0,0.1,0.2\}$ . There is a curious effect, that for N$\neq$0, as we
increase $\varepsilon,$ the death time first decreases and subsequently it
behaves "normally", by increasing with $\varepsilon$. In the Fig. \ref{t6} we
show the revival time as a function of $\varepsilon$ for the same values of
$N$. In all cases the revival time decreases with $\varepsilon$.%
\begin{figure}
[ptbhptbh]
\begin{center}
\includegraphics[natheight=2.674000in,
natwidth=5.366200in, height=2.3367in, width=2.3367in
]%
{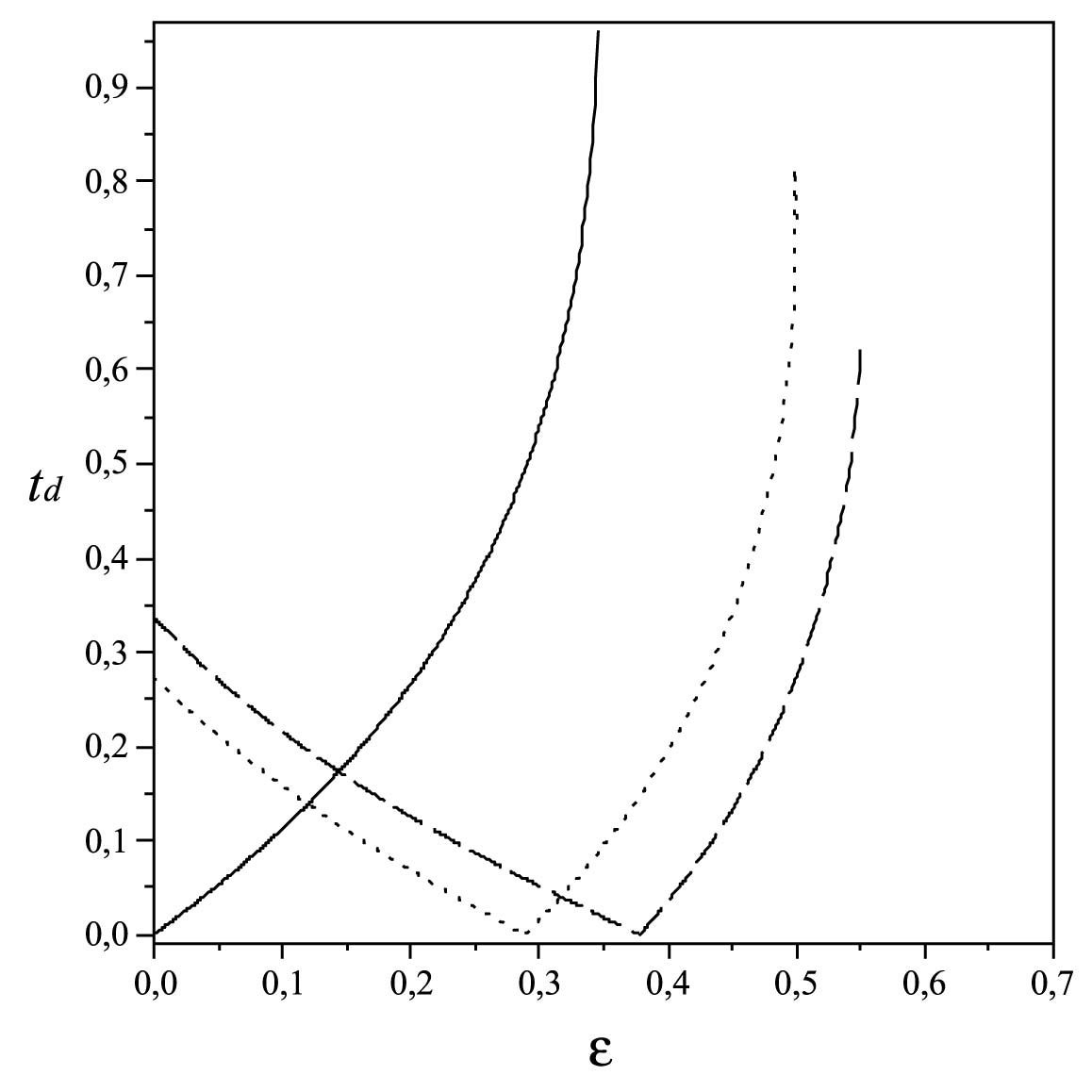}%
\caption{Death time with initial $|\Psi_{1}\rangle$ and :$N=0$ (solid line),
$N=0.1$ (dotted line), $N=0.2$ (dashed line)}%
\label{t5}%
\end{center}
\end{figure}
\begin{figure}
[ptbhptbhptbh]
\begin{center}
\includegraphics[natheight=2.674000in,
natwidth=5.366200in, height=2.3367in, width=2.3367in
]%
{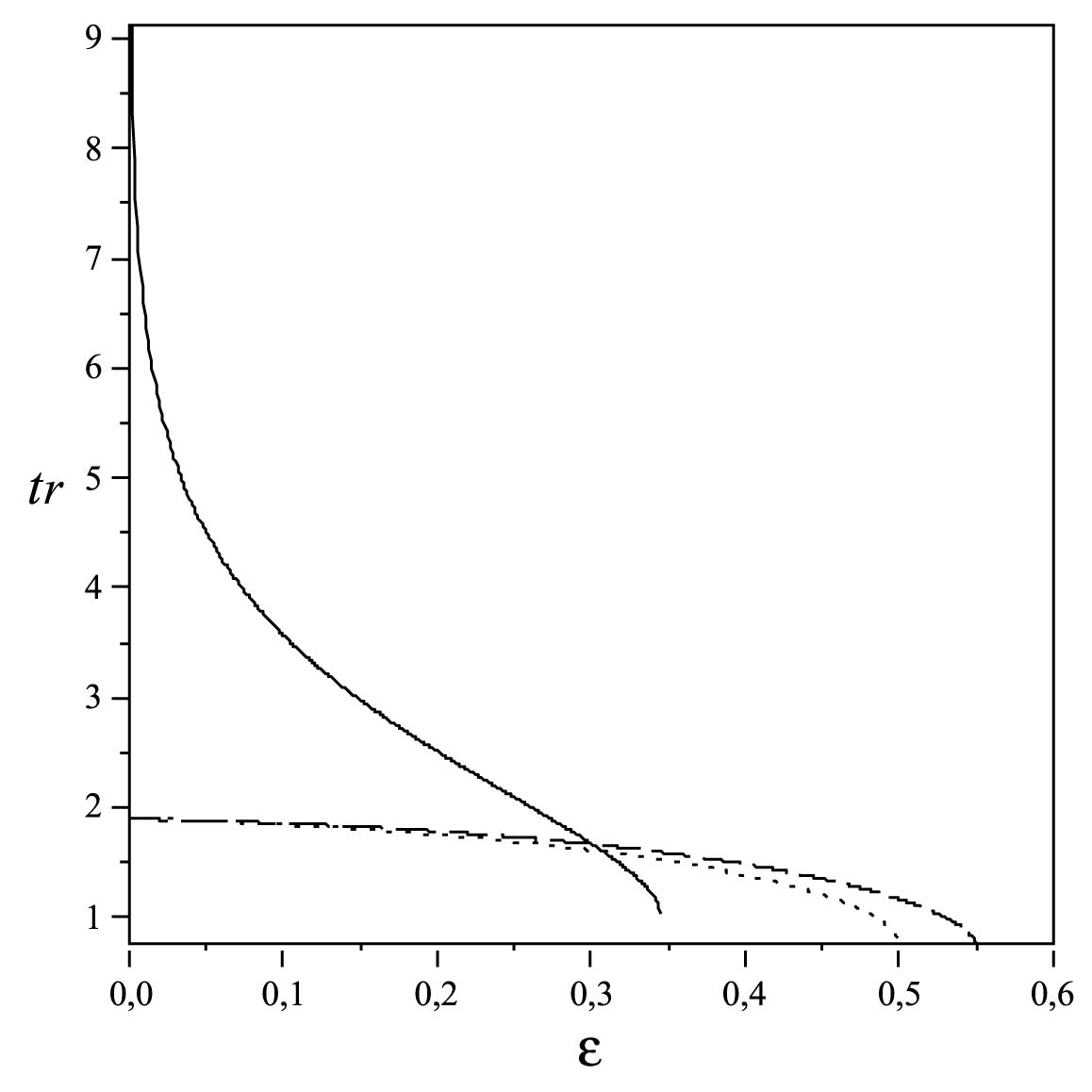}%
\caption{Revival time with initial $|\Psi_{1}\rangle$ and :$N=0$ (solid line),
$N=0.1$ (dotted line), $N=0.2$ (dashed line)}%
\label{t6}%
\end{center}
\end{figure}

\item[d)] Finally, we consider the case with initial $|\Psi_{2}(t)\rangle
=\varepsilon|\phi_{2}\rangle+\sqrt{1-\varepsilon^{2}}|\phi_{3}\rangle$ . Its
concurrence is: $C(\rho(t))=\max\{0,C1(t),C2(t)\}$, with $C1(t)$ and $C2(t)$
defined in (\ref{C21},\ref{C22}), and its initial value:$C(\Psi_{2}%
(0))=|2\varepsilon^{2}-1|$. Fig. \ref{c7}. shows the time evolution of the
concurrence with $N=0.1$ for several values of \ $\varepsilon$ .%
\begin{figure}
[ptbh]
\begin{center}
\includegraphics[natheight=2.674000in,
natwidth=5.366200in, height=2.2451in, width=2.437in
]%
{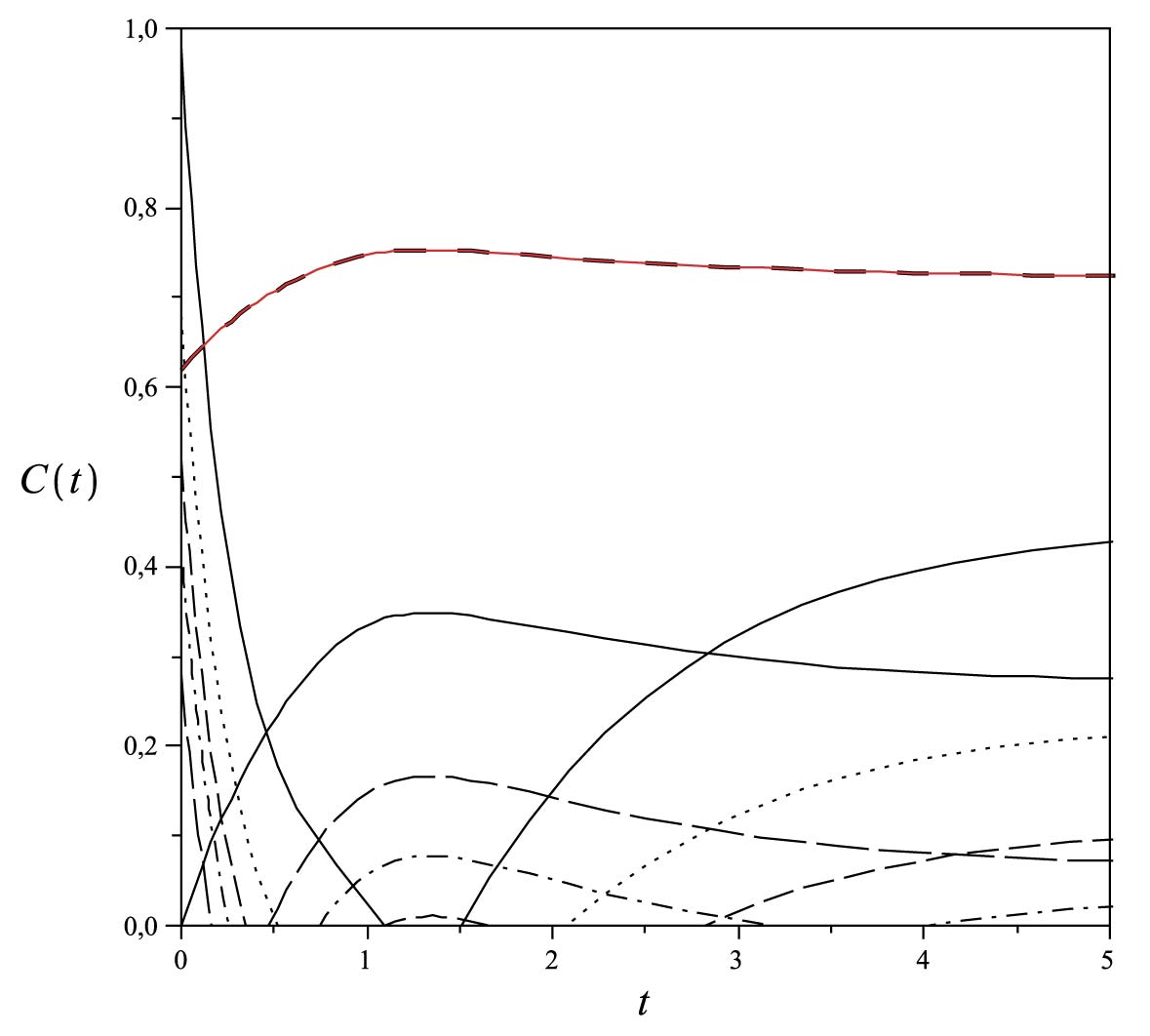}%
\caption{Time evolution of concurrence for initial $|\Psi_{2}\rangle$ and:
$\varepsilon=0.1$ (solid line), $\varepsilon=0.4$ (dotted line),
$\varepsilon=0.49$ (dashed line), $\varepsilon=0.54$ (dash dotted line),
$\varepsilon=0.6$ (long dashed line), $\varepsilon=0.9$ ( space dashed line),}%
\label{c7}%
\end{center}
\end{figure}

As we can see from the Fig. \ref{c7}, this case is more complex, since there
are more than one death and revival before reaching the critical value of
$\varepsilon$. Such a situation has been described previously \cite{fi,e4}.
Like in the previous cases above a certain critical $\varepsilon$, when we get
"close" to the DFS, these effects disappear and $C(t)$ goes asymptotically to
its stationary value.
\end{enumerate}

\bigskip

\section{Discussion}

The first and most obvious observation is that if we start with an initial
state that is in the DFS plane, the local and non-local coherences are not
affected by the environment, thus it experiences no decoherence and the
concurrence is constant in time. It does increase with the squeeze parameter N
(in the case of initial $|\phi_{1}\rangle$), getting maximum entanglement for
N$\rightarrow\infty.$ So this reservoir is not acting as a thermal one, in the
sense that introduces $\operatorname{rand}$omness. On the contrary, a common
squeezed bath tends to enhance the entanglement, as we increase the parameter N.

This is clear if we observe that for N$\rightarrow\infty$, $|\phi_{1}%
\rangle\rightarrow\frac{1}{\sqrt{2}}(|++\rangle+|--\rangle)$, which is a Bell
state. On the other hand, if we start with the initial state $|\phi_{2}%
\rangle$, this state is independent of N and it is also maximally entangled,
so C=1 for all times and all N%
\'{}%
s.

Now, we consider other situations with initial states outside the DFS:

For N=0, $|\phi_{1}\rangle\rightarrow|--\rangle;|\phi_{4}\rangle
\rightarrow|++\rangle$ while $|\phi_{2}\rangle$ and $|\phi_{3}\rangle$ are
independent of N and represent the antisymmetrical singlet and one of the
triplet states, respectively.

If our initial state is $|\phi_{3}\rangle$, the initial concurrence is
C(t=0)=1, and for the steady state, C(t=$\infty$)=0. On the other hand, from
the negativity of the eigenvalues of $\rho^{PT}$ (Peres-Horodecki criteria),
we know that the state is entangled for all times, so presumably \ the
concurrence versus time is a smooth curve starting from 1 and going
asymptotically to zero.

For an initial $|\phi_{4}\rangle=|++\rangle$, of course C(t=0)=0, since we
have a factorized state. Since this initial state is symmetrical, the steady
state is $|\phi_{1}\rangle=|--\rangle$, which is also a factorized state with
C(t=$\infty$)=0. Our analysis shows that C(t)=0 for all times.

Finally, if we have initial states of the form $|\Psi_{1}\rangle
=\varepsilon|\phi_{1}\rangle+\sqrt{1-\varepsilon^{2}}|\phi_{4}\rangle
=\varepsilon$ $|--\rangle+\sqrt{1-\varepsilon^{2}}$ $|++\rangle$ or $|\Psi
_{2}\rangle=\varepsilon|\phi_{2}\rangle+\sqrt{1-\varepsilon^{2}}|\phi
_{3}\rangle=\frac{1}{\sqrt{2}}[(\varepsilon+\sqrt{1-\varepsilon^{2}})$
$|-+\rangle-(\varepsilon-\sqrt{1-\varepsilon^{2}})$ $|+-\rangle]$, there is
sudden death if $0<\varepsilon<0.34525$ and $0<\varepsilon<0.707$
respectively. Obviously, when we get "near" the DFS, that is $\varepsilon$
gets larger than zero, the sudden death times become larger up to some
critical value, for which \ the death time becomes infinite and we no longer
observe the sudden death. Thus, we have related directly the local decoherence
with the disentanglement.

Another interesting feature is the revival. For the $|\Psi_{1}\rangle$ case,
after a finite time period, the entanglement "revives", and the concurrence
reaches asymptotically its steady state value. This time gap becomes smaller
for larger $\varepsilon$ until we reach $\varepsilon=0.34525$ . At this point,
the two times merge and beyond that, there is no longer sudden death nor revival.

On the other hand, for the $|\Psi_{2}\rangle$ case, the sudden death and
revival happen simultaneously, thus the above mentioned time period vanishes.
The phenomena of one or periodical revivals have been obtained before, but
always in the context of one single reservoir connecting both atoms, like in
the present case \cite{fi,fi2,e4}.

\bigskip In the N$\neq0$ case$,$ we also observe sudden death and revivals, up
to a certain "distance" to the DFS ( or more precisely, up to a certain
critical value of $\varepsilon$).

However, there are certain differences with the N=0 case.

For the initial state $|\phi_{4}\rangle$, the death time versus N first
increases for small values of N, and for N$\gtrsim0.421$, it tends to
decrease. A possible interpretation of the increase is the following one:

\bigskip$|\phi_{4}\rangle$=$\frac{1}{\sqrt{N^{2}+M^{2}}}(M|++\rangle
-N|--\rangle)=\frac{N}{\sqrt{N^{2}+M^{2}}}(\sqrt{1+\frac{1}{N}}|++\rangle
-|--\rangle)$, so that the ratio of the probabilities of the double excited
and the ground states goes as $(1+\frac{1}{N}.)$

On the other hand, the squeezed vacuum has only components for the even number
of photons, so the interaction between our system and the reservoir goes by
pairs of photons. Now, for very small N, the average photon number is also
small, so the predominant interaction with the reservoir will be the doubly
excited state that would tend to decay via two photon spontaneous emission..
Now in the $|\phi_{4}\rangle$ case, the population of the $|++\rangle$ goes
down with N, meaning that the interaction with the reservoir goes also down
with N and therefore, the death time will necessarily increase with N, which
describes qualitatively the first part of the curve (fig \ref{t4}-a). On the
other hand, as we increase the average photon number N, other processes like
the two photon absorption will be favored, and since there will be more
photons and the $|--\rangle$ \ population tends to increase with N, this will
enhance the system-bath interaction and therefore the death of the
entanglement will occur faster, or the death time will decrease.

In the $|\phi_{3}\rangle$ case, initially there is no $|++\rangle$ component,
thus we expect a higher initial death time. However this case is different
from the previous one in the sense that the state is independent of N, so
there is no initial increase. However, as the state evolves in time,
$|++\rangle$ and $|--\rangle$ components will build up and the argument for
the decrease of the death time with N follows the same logic as in the
previous case.(fig \ref{t3}-a).

\subsection{Acknowledgements}

MH was supported by a Conicyt grant.

M.O was supported by Fondecyt \# 1051062.

The authors thank Prof. Sascha Wallentowitz for useful discussions.

\smallskip

\bigskip

\section{APPENDIX}

In the representation spanned by $\{|\phi_{1}\rangle,|\phi_{2}\rangle
,|\phi_{3}\rangle,|\phi_{4}\rangle\}$, the solution of master equation
(\ref{em}) is:

\begin{appendices}
\begin{align*}
\rho_{11}(t)  & =\frac{4}{\sqrt{N(N+1)}(8N+4)}\{[-(2N+1)(\rho_{44}%
(0)+\rho_{33}(0))\\
& \times\sqrt{\frac{N(N+1)}{4}}+(\rho_{44}(0)+\rho_{33}(0)))N^{2}\\
& +(\rho_{44}(0)+\rho_{33}(0))N+\frac{1}{4}\rho_{44}(0)]e^{-2t(\sqrt{N}%
+\sqrt{N+1})^{2}}\\
& +[-(2N+1)(\rho_{44}(0)+\rho_{33}(0))\sqrt{\frac{N(N+1)}{4}}\\
& -(\rho_{44}(0)+\rho_{33}(0)))N^{2}-(\rho_{44}(0)+\rho_{33}(0))N\\
& -\frac{1}{4}\rho_{44}(0)]e^{-2t(\sqrt{N}-\sqrt{N+1})^{2}}\\
& +(2N+1)(\rho_{44}(0)+\rho_{33}(0)+\rho_{11}(0))\sqrt{N(N+1)}\}
\end{align*}
\[
\rho_{12}(t)=\rho_{12}(0)
\]%
\begin{align*}
\rho_{13}(t)  & =\frac{12e^{-(2N+1)t}}{\sqrt{N}(24N^{3}+36N^{2}+10N-1)}%
[-\frac{2}{3}(N+\frac{1}{2})\\
& \times(N\sqrt{N+1}+\frac{1}{4}\sqrt{2N+1}\sqrt{2N^{2}+N})\\
& \times(\rho_{43}(0)-e^{i\Psi}\rho_{34}(0))e^{t\frac{(-\sqrt{2N^{2}%
+N}(2N+1)+4N\sqrt{N+1}\sqrt{2N+1})}{\sqrt{2N^{2}+N}}}\\
&
-\frac{2}{3}(N+\frac{1}{2})(N\sqrt{N+1}-\frac{1}{4}\sqrt{2N+1}\sqrt
{2N^{2}+N})\\
& \times(e^{i\Psi}\rho_{34}(0)+\rho_{43}(0))e^{-t\frac{(\sqrt{2N^{2}%
+N}(2N+1)+4N\sqrt{N+1}\sqrt{2N+1})}{\sqrt{2N^{2}+N}}}\\
& +(-\frac{1}{3}(N+\frac{1}{2})\rho_{34}(0)e^{i\Psi}+\rho_{13}(0)(N^{2}%
+N-\frac{1}{12}))\\
& \times\sqrt{2N+1}\sqrt{2N^{2}+N}+\frac{4}{3}(N+\frac{1}{2})\sqrt{N+1}%
\rho_{43}(0)]
\end{align*}
\qquad%
\begin{align*}
\rho_{14}(t)  &
=\frac{8e^{-(2N+1)t}}{(2N+1)(12N^{2}+12N-1)}[e^{-(2N+1+4\sqrt
{N(N+1)})t}\\
&
\times(-\frac{1}{2}(N+\frac{1}{2})(2\rho_{44}(0)+\rho_{33}(0))\sqrt
{N(N+1)}\\
& +(\frac{1}{2}\rho_{44}(0)+\rho_{33}(0))(N^{2}+N)+\frac{1}{8}\rho_{44}(0))\\
& -e^{-(2N+1-4\sqrt{N(N+1)})t}(\frac{1}{2}(N+\frac{1}{2})\\
& \times(2\rho_{44}(0)+\rho_{33}(0))\sqrt{N(N+1)}\\
& +(\frac{1}{2}\rho_{44}(0)+\rho_{33}(0))(N^{2}+N)+\frac{1}{8}\rho_{44}(0))\\
& +\frac{3}{2}(\sqrt{2N+1}\rho_{14}(0)(N^{2}+N-\frac{1}{12})\sqrt{2N^{2}+N}\\
&
+\frac{2}{3}(N+\frac{1}{2})(2\rho_{44}(0)+\rho_{33}(0))N\sqrt{N+1})]
\end{align*}
\[
\rho_{21}(t)=\rho_{21}(0)
\]
\[
\rho_{22}(t)=\rho_{22}(0)
\]
\[
\rho_{23}(t)=\rho_{23}(0)e^{-(2N+1)t}%
\]
\[
\rho_{24}(t)=\rho_{24}(0)e^{-(2N+1)t}%
\]
\begin{align*}
\rho_{31}(t)  & =\frac{-8e^{-i\Psi}e^{-(2N+1)t}}{\sqrt{N}(24N^{3}%
+36N^{2}+10N-1)}[(N\sqrt{N+1}{}\\
& {}-\frac{1}{4}\sqrt{2N+1}\sqrt{2N^{2}+N})(N+\frac{1}{2}){}\\
&
{}\times(e^{i\Psi}\rho_{34}(0)+\rho_{43}(0))e^{\frac{-(((2N+1)t+I\Psi
)\sqrt{2N^{2}+N}+4tN\sqrt{N+1}\sqrt{2N+1}}{\sqrt{2N^{2}+N}}}{}\\
&
{}+e^{i\Psi}(N+\frac{1}{2})(N\sqrt{N+1}+\frac{1}{4}\sqrt{2N+1}\sqrt
{2N^{2}+N}){}\\
&
{}\times(e^{i\Psi}\rho_{34}(0)-\rho_{43}(0))e^{\frac{-(((2N+1)t+I\Psi
)\sqrt{2N^{2}+N}-4tN\sqrt{N+1}\sqrt{2N+1}}{\sqrt{2N^{2}+N}}}{}\\
& {}-\frac{2}{3}\sqrt{2N+1}\sqrt{2N^{2}+N}(\rho_{31}(0)e^{i\Psi}(N^{2}%
+N-\frac{1}{2}){}\\
& {}-\frac{1}{3}(N+\frac{1}{2})\rho_{43}(0))-(2N+1)N\sqrt{N+1}\rho
_{34}(0)e^{i\Psi}]
\end{align*}
\newline%
\[
\rho_{32}(t)=\rho_{32}(0)e^{-(2N+1)t}%
\]
\begin{align*}
\rho_{33}(t)  & =\frac{1}{2}[e^{-2t(\sqrt{N}-\sqrt{N+1})^{2}}(\rho
_{33}(0)+\rho_{44}(0)\frac{(N+1)}{\sqrt{N(N+1)}})\\
& +[e^{-2t(\sqrt{N}+\sqrt{N+1})^{2}}(\rho_{33}(0)-\rho_{44}(0)\frac
{(N+1)}{\sqrt{N(N+1)}}]
\end{align*}
\begin{align*}
\rho_{34}(t)  &
=\frac{1}{2}(\rho_{34}(0)-e^{-i\Psi}\rho_{43}(0))e^{-2t(\sqrt
{N}-\sqrt{N+1})^{2}}\\
&
+\frac{1}{2}(\rho_{34}(0)+e^{-i\Psi}\rho_{43}(0))e^{-2t(\sqrt{N}+\sqrt
{N+1})^{2}}%
\end{align*}
\begin{align*}
\rho_{41}(t)  & =\frac{12e^{-(2N+1)t}}{N\sqrt{N+1}(2N+1)(12N^{2}+12N-1)}%
[\frac{1}{3}N\sqrt{N+1}\\
& \times(-(2N+1)(\frac{\rho_{33(0)}}{2}+\rho_{44}(0))\sqrt{N(N+1)}\\
& +(2\rho_{33}(0)+\rho_{44}(0))(N^{2}+N)\\
&
+\frac{\rho_{44}(0)}{4})e^{-(2N+1+4\sqrt{N(N+1)})t}-\frac{1}{3}N\sqrt
{N+1}\\
& \times((2N+1)(\frac{\rho_{33(0)}}{2}+\rho_{44}(0))\sqrt{N(N+1)}\\
& +(2\rho_{33}(0)+\rho_{44}(0))(N^{2}+N)\\
& +\frac{\rho_{44}(0)}{4})Ne^{-(2N+1-4\sqrt{N(N+1)})t}\\
& +\sqrt{N(N+1)}(\sqrt{2N+1}\rho_{41}(0)(N^{2}-\frac{1}{12}+N)\\
& \times\sqrt{2N^{2}+N}+\frac{4}{3}(N+\frac{1}{2})N\sqrt{N+1}N\\
& \times(\frac{\rho_{33}(0)}{2}+\rho_{44}(0)))]
\end{align*}
\[
\rho_{42}(t)=\rho_{42}(0)e^{-(2N+1)t}%
\]
\begin{align*}
\rho_{43}(t)  &
=\frac{1}{2}(\rho_{43}(0)-e^{i\Psi}\rho_{34}(0))e^{-2t(\sqrt
{N}-\sqrt{N+1})^{2}}\\
&
+\frac{1}{2}(\rho_{43}(0)+e^{i\Psi}\rho_{34}(0))e^{-2t(\sqrt{N}+\sqrt
{N+1})^{2}})
\end{align*}
\newline%
\begin{align*}
\rho_{44}(t)  &
=(\frac{\rho_{44}(0)}{2}-\frac{\sqrt{N(N+1)}}{2N+1}\rho
_{33}(0))e^{-2t(\sqrt{N}+\sqrt{N+1})^{2}}\\
&
+(\frac{\rho_{44}(0)}{2}+\frac{\sqrt{N(N+1)}}{2N+1}\rho_{33}(0))e^{-2t(\sqrt
{N}-\sqrt{N+1})^{2}}%
\end{align*}
\end{appendices}

\end{document}